\documentclass{article}
\pdfoutput=1
\usepackage{amsmath}
\usepackage{amssymb}
\usepackage{graphicx}
\usepackage{color}
\usepackage[usenames,dvipsnames,svgnames,table]{xcolor}
\usepackage{soul}
\usepackage{subfig}
\usepackage{ulem}
\usepackage{float}
\usepackage{multirow}

\pdfpageheight\paperheight
\pdfpagewidth\paperwidth

\newcommand{\avg}[1]{\langle #1 \rangle}

\usepackage{jcappub}

\begin{document}
\title{Analytical covariance between voxel intensity distributions and line-intensity mapping power spectra}
\author[a]{Gabriela Sato-Polito}
\author[a]{José Luis Bernal}

\affiliation[a]{William H. Miller III Department of Physics and Astronomy, Johns Hopkins University, 3400 North Charles Street, Baltimore, MD 21218, United States}

\emailAdd{gsatopo1@jhu.edu, jbernal2@jhu.edu}

\abstract{The power spectrum and the voxel intensity distribution (VID) are two of the main proposed summary statistics to study line-intensity maps. We reformulate the derivation of the VID in terms of the local overdensities and derive for the first time an analytic covariance between the VID and the line-intensity mapping power spectrum. We study the features of this covariance for different experimental setups and show that we can recover similar results to simulation-based covariances. With this formalism, we also compute the cosmic variance contribution the VID uncertainty, which we find to be subdominant with respect to the standard variance from Poisson sampling. Our results allow for general joint analyses of the VID and the line-intensity mapping power spectrum.}

\maketitle

\hypersetup{pageanchor=true}

\section{Introduction}
Line-intensity mapping (LIM) has recently emerged as a promising technique to survey large cosmological volumes that has the potential to recover precise redshift information by targeting a variety of bright spectral lines~\cite{Kovetz_report}. Since the resulting maps capture the integrated specific intensity, these measurements are sensitive to both cosmology and astrophysics. The former is encoded in the clustering of galaxies and gas in the intergalactic medium that emit the desired signal, and the later, in the luminosity produced by each emitter. 

The most well-studied line is the 21-cm spin-flip transition in neutral hydrogen (see, e.g., Ref.~\cite{2010ARA&A..48..127M}) and there are a variety of ongoing or planned experiments that target cosmic dawn and the epoch of reionization~\cite{Ali:2015uua, 2013A&A...556A...2V, Tingay:2012ps, 2013ApJ...763L..20M, DeBoer:2016tnn, Koopmans:2015sua} and the late Universe~\cite{Bandura:2014gwa, Newburgh:2016mwi, MeerKLASS:2017vgf}. In addition, several other line transitions at higher frequencies have garnered attention as probes of the different phases of the interstellar medium, such as the rotational lines of carbon monoxide (CO)~\cite{2011ApJ...741...70L, Breysse:2014uia, Li:2015gqa, 2013ApJ...768...15P, Padmanabhan:2017ate}, the fine structure line of ionized carbon ([CII])~\cite{Silva:2014ira, Pullen:2017ogs}, H$\alpha$ and H$\beta$~\cite{2017ApJ...835..273G, 2017arXiv171109902S}, Lyman-$\alpha$~\cite{Pullen:2013dir, 2013ApJ...763..132S}, and oxygen lines~\cite{2017ApJ...835..273G}. A vast number of experiments that target these spectral lines are already observing or currently under development (see e.g.,~\cite{Cleary_COMAP,CCAT-prime,EXCLAIM,TIM, SPHEREx,HETDEX}), with some initial constraints and tentative power spectrum detections~\cite{Keating:2015qva, Keating:2016pka, Croft_detection, Pullen:2017ogs, Yang_CII, Keenan:2021uue, Ihle:2021fkp}.

Different methods have been proposed to extract information from line-intensity maps. While the LIM power spectrum is the most prominent statistic, the cosmological information it contains is inherently limited and degenerate with astrophysical uncertainties~\cite{Padmanabhan_astrocosmo, Bernal_guide}. Complementary probes have been proposed to access non-Gaussian information and break degeneracies with astrophysical parameters. Among them, the voxel intensity distribution (VID), which is an estimator of the 1-point probability distribution function of the temperature measured in a voxel, has been shown to be very promising~\cite{Breysse:2015saa, Breysse:2016szq}. The VID depends directly on the line-luminosity function and is more sensitive to the high-luminosity end that peaks above the instrumental noise, whereas the power spectrum is more sensitive to cosmological parameters and depends on a weighted average of luminosities that favors comparatively fainter objects.

While both the power spectrum and the VID have individually been shown to offer promising constraints to astrophysical and cosmological parameters, as well as extensions to $\Lambda$CDM (see e.g.,~ \cite{Bernal_limdm,Bernal_limnu,Azadeh_fNL, Bernal_baolim, Liu:2020izx, MoradinezhadDizgah:2021upg, Bauer:2020zsj}), the more substantial gain lies in their combination. The information contained in the LIM power spectrum and the VID are highly complimentary, and a joint analysis has been shown to significantly increase the precision of the inference of the line-luminosity function~\cite{COMAP:2018kem}.  A joint analysis requires a covariance matrix to properly account for the shared information content of the observables. While Ref.~\cite{COMAP:2018kem} relied on simulations to estimate their covariance empirically, a theoretical understanding of the expected covariance between these observables is still lacking.


We derive, for the first time, an analytic covariance between the VID and the LIM power spectrum for spectral 
lines emitted within halos. We first reformulate the standard derivation of the VID signal by an equivalent expression that explicitly accounts for the local perturbations of the number of emitters --- a biased tracer of the local matter overdensities. This enables us to include effect of cosmic variance --- the variance of the brightness temperature field on the scales of the size of the survey --- as a contribution to the predicted statistical error of the VID, though we argue that it is generally a negligible contribution. Using the same reformulation of the VID, we compute its covariance with the LIM power spectrum. We focus on the Legendre monopole of the LIM power spectrum and in observations performed using only the autocorrelation between antennas, but this work can be straightforwardly extended for higher order multipoles of the power spectrum or interferometers. 

The derivation presented in this work shows that we can interpret the covariance between the VID and the power spectrum as the response of the power spectrum to the mean density perturbation. Since the VID depends (among other things) on the perturbation in the number density of emitters $\delta_h$ and the power spectrum depends on two powers of the temperature fluctuation $\delta T$, the resulting covariance is proportional to the bispectrum $ \avg{\delta_h\delta T\delta T}$ integrated over two of the wavenumbers. We note that the type of integrated bispectrum we derive here has been previously studied in the context of position-dependent power spectra and matter density PDF~\cite{Chiang:2014oga, Jamieson:2020wxf}, derived using the separate universe approach.


This paper is structured as follows. First, we introduce the theoretical modeling of the VID, the power spectrum and all the required quantities to compute their covariance in Sec.~\ref{sec:theory}. We discuss the reformulation of the probability distribution function of the temperature measured in a voxel as function of the local density of emitters in Sec.~\ref{sec:PDF} and used it to derive the cosmic variance contribution to the VID variance and the covariance between the VID and the power spectrum in Sec.~\ref{sec:covariance}. Finally, we conclude in Sec.~\ref{sec:conclusions}. We include a derivation of the shot noise bispectrum in App.~\ref{app:shot}, details on the quantities required to compute the power spectrum and bispectrum in App.~\ref{app:PT} and compare the analytic covariance derived in this work with the numerical results of Ref.~\cite{COMAP:2018kem} in App.~\ref{sec:comparison}.

Throughout this work we consider the best-fit cosmological parameters from the full data set of \textit{Planck} assuming $\Lambda$CDM~\cite{Planck:2018vyg} and adopt the following Fourier transform convention:
\begin{equation}
    f(\pmb{k}) = \int {\rm d}^3\pmb{x} \ f(\pmb{x})e^{-i\pmb{kx}}, \qquad f(\pmb{x}) = \int \frac{{\rm d}^3\pmb{k}}{(2\pi)^3}\ f(\pmb{k})e^{i\pmb{kx}}.
\end{equation}

\section{Theoretical Modeling}
\label{sec:theory}
In this section we briefly review the modeling of the VID and the power spectrum, as well as the quantities that will be necessary for the calculation of the variance of the VID and the covariance between the VID and the power spectrum. We begin by assuming that the measured line emission is sourced within dark matter halos, which trace the underlying matter distribution, and relate the line luminosity to halo mass. 

The brightness temperature $T$ at a position $\boldsymbol{x}$ of a given emission line with rest-frame frequency $\nu$ is related to its local luminosity density $\rho_{L}$ as
\begin{equation}
    T(\boldsymbol{x}) = \frac{c^3(1+z)^2}{8\pi k_B \nu^3 H(z)}\rho_{L}(\boldsymbol{x}) \equiv X_{\rm LT}\rho_L(\boldsymbol{x})\,,
\end{equation}
where $c$ is the speed of light, $z$ is the redshift of the emission, $k_B$ is the Boltzmann constant, $H$ is the Hubble parameter and we have defined $X_{\text{LT}}$ in the second equality to compress our expressions. The luminosity density can be computed assuming a relation between the specific luminosity $L$ and the halo mass $M$. Thus, we can write the $m$-th moment of the temperature distribution as 
\begin{equation}
    \langle T^m\rangle = X_{\rm LT}^m\exp\left\lbrace\left(m^2-m\right)\frac{\sigma_{L}^2\log^2 10}{2}\right\rbrace\int{\rm d}ML^m(M)\frac{{\rm d}n}{{\rm d}M}\,,
    \label{eq:Tmoments}
\end{equation}
where ${\rm d}n/{\rm d}M$ is the halo mass function and $\sigma_L$ is a mean-preserving log-scatter in the luminosity-halo mass relation. 

\subsection{VID}
The VID is the histogram of temperatures measured within the observed voxels; we follow Ref.~\cite{Breysse:2016szq} to compute it. A voxel of volume $V_{\rm vox}$ that contains $N$ emitters will, in the absence of noise, have a brightness temperature of 
\begin{equation}
    T  
    = \frac{X_{\text{LT}}}{V_{\text{vox}}} \sum_{i=1}^{N} L_i,
    \label{eq:T}
\end{equation}
where $L_i$ is the luminosity of the $i$-th emitter. 
%
Hence, the probability of observing a temperature $T$ for a voxel that contains only a single emitter is given by
\begin{equation}
    \mathcal{P}_1 (T) = \frac{V_{\text{vox}}}{\bar{n}X_{\text{LT}}} \frac{{\rm d}n}{{\rm d}L}\bigg{|}_{L=TV_{\text{vox}}/X_{\text{LT}}},
\end{equation}
where $\bar{n}$ is the mean comoving number density of emitters and ${\rm d}n/{\rm d}L$ is the luminosity function of the targeted line, which we compute from the halo mass function and the $L(M)$ relation. Since the contribution of each emitter to the observed temperature is additive, the probability of observing a given temperature in a voxel with $N$ emitters is $\mathcal{P}_1$ convolved $N$ times with itself. Similarly, in voxels where there are no emitters we have  $\mathcal{P}_0(T)=\delta_D(T)$, where $\delta_D$ is the Dirac delta. 

The total probability distribution function (PDF) $\mathcal{P} (T)$ of observing a voxel with temperature $T$ is therefore given by the probability $\mathcal{P}(N)$ that a voxel contains $N$ sources and that these $N$ sources produce an aggregate temperature $T$, summed over all possible values of $N$. That is,
\begin{equation}
    \mathcal{P} (T) = \sum_{N=0}^{\infty} \mathcal{P}_{N}(T) \mathcal{P}(N).
    \label{eq:PT}
\end{equation}

If the sources are unclustered, $\mathcal{P}(N)$ is a Poisson distribution $\mathcal{P}_{\text{Poiss}}(N,\bar{N}_h)$, where the mean number $\bar{N}_h$ of emitters in a voxel is determined by the mean number density of dark matter halos in the universe, $\bar{N}_h = \bar{n}_h V_{\text{vox}}$. To include the effect of clustering in the calculation of $\mathcal{P}(N)$, Ref.~\cite{Breysse:2016szq} uses the fact that the halo number-count follows the matter distribution, which can be approximated by a lognormal~\cite{Coles_lognormal}. The number of emitters is then a Poisson draw where the mean is now determined by the expected number of (lognormally distributed) emitters in a given voxel. Under these assumptions, we have
\begin{equation}
    \mathcal{P}(N)=\int {\rm d}\eta \mathcal{P}_{\rm LN}(\eta)\mathcal{P}_{\rm Poiss}(N,\eta)\,,
\label{eq:PN_trad}    
\end{equation}
where $\eta$ refers to the expected number of emitters and $\mathcal{P}_{\rm LN}$ denotes a lognormal distribution for $\eta$, assuming that the halo number count fluctuations can be expressed in terms of a Gaussian random variable and  its variance~\cite{Kayo_LN}.

Finally, the total observed temperature includes also the thermal noise (in addition to the contribution of continuum foregrounds and line interlopers, which we neglect in this work). Therefore, the total probability distribution function is
\begin{equation}
    \mathcal{P}_{\rm tot}(T) = \left(\mathcal{P}_{\rm noise}*\mathcal{P}\right)(T)\,,
    \label{eq:VIDnoise}
\end{equation}
where usually $\mathcal{P}_{\rm noise}$ is a Gaussian distribution with standard deviation given by the effective instrumental noise per voxel and $*$ denotes the convolution operator. 

The probability of observing a voxel temperature within a given range $\Delta T_i \equiv \left( T_i^{\rm min}, T_i^{\rm max}\right]$ can be estimated with the VID $\mathcal{B}_i$, which is the total number of voxels with an observed temperature within $\Delta T_i$. The relation between $\mathcal{B}_i$, which is the actual observable, and $\mathcal{P}(T)$ is
\begin{equation}
    \mathcal{B}_i = N_{\rm vox}\int_{\Delta T_i}{\rm d}T\mathcal{P}_{\rm tot}(T)\,,
    \label{eq:Balpha}
\end{equation}
where $N_{\rm vox}=V_{{\rm field}}/V_{\rm vox}$ is the total number of voxels in an observed volume $V_{{\rm field}}$.


\subsection{Correlations}\label{sec:correlations}
The total observed LIM power spectrum consists of three components: clustering, shot noise, and instrument noise, which we can write as
\begin{equation}
    \tilde{P}^{\rm tot}_{TT}(k, \mu, z) = \tilde{P}^{\rm clust}_{TT}(k,\mu, z) + \tilde{P}^{\rm shot}_{TT}(k, \mu, z) + P^{\rm N}(z),
\end{equation}
where the tilde denotes an observed quantity and $\mu$ is the cosine of the angle between the wave number vector ${\pmb k}$ and the line of sight. The observed power spectrum differs from the predicted one due to the limited experimental resolution and the volume probed; this is why the observed contribution from shot noise also depends on $k$ and $\mu$. Hence, we include the window functions that model these limitations in the observed power spectrum, such that, for any power spectrum, 
\begin{equation}
    \tilde{P}(k,\mu)=\int \frac{{\rm d^3}\boldsymbol{q}}{(2\pi)^3}W^2_{\rm vol}(\boldsymbol{k})W_{\rm vox}^2(\boldsymbol{q}-\boldsymbol{k})P(\boldsymbol{q}-\boldsymbol{k}),
    \label{eq:obsPk}
\end{equation}
where $W_{\rm vol}$ models the limited survey volume and $W_{\rm vox}$ the voxel resolution. We will discuss the particular forms chosen for these functions in more detail in Sec.~\ref{sec:experiment}. However, we note that, since the volume window simply includes or excludes certain spatial positions, it corresponds to a product with the density field in configuration space and therefore a convolution in Fourier space. Conversely, the voxel window captures the loss of information on scales smaller than the voxel size and therefore this smoothing is added as a convolution in configuration space and as a product in Fourier space.

Since the line emission is sourced within dark matter halos, the temperature fluctuations are a biased tracer of the matter distribution. We consider a simple linear bias model to describe the relation between temperature and matter perturbations and assume Poisson shot noise. We therefore model the LIM power spectrum as
\begin{equation}
    P_{TT}(k,\mu, z) \equiv P^{\rm clust}_{TT}(k,\mu, z) + P_{TT}^{\rm shot}(z) =  \avg{T}^2 \left(b^T_1 + f\mu \right)^2  P_l + \avg{T^2} 
    \label{eq:PTT}
\end{equation}
where $P_l$ is the linear power spectrum of cold dark matter and baryons, $b_1^T$ is the linear bias parameter of the temperature field, and $f$ is the linear scale-independent growth rate. We discuss the noise power spectrum $P_N$ in Sec.~\ref{sec:experiment}.

The key ingredient for the calculation of the covariance between the VID and the power spectrum is a bispectrum of the form $\avg{\delta_h\delta T\delta T}$. This bispectrum includes contributions from clustering and shot-noise, but does not have any contribution from instrumental noise since the temperature fluctuations due to thermal noise are uncorrelated with the halo density. The total bispectrum is given by
\begin{equation}
    B_{hTT}(\pmb{k}_1, \pmb{k}_2,\pmb{k}_3) = B^{\rm clust}_{hTT}(\pmb{k}_1, \pmb{k}_2,\pmb{k}_3) + \frac{P_{TT}^{\rm clust}(\pmb{k}_2)+P_{TT}^{\rm clust}(\pmb{k}_3)}{\bar{n}}+\langle T^2\rangle\left(\frac{P_{hT}^{\rm clust}(\pmb{k}_1)}{\langle T\rangle}+\frac{1}{\bar{n}}\right)\,.
    \label{eq:BhTT}
\end{equation}
The last two terms of the bispectrum correspond to shot noise contributions, which we assume to be Poissonian; we include their derivation in App.~\ref{app:shot}. Using standard perturbation theory (see, e.g., Ref.~\cite{Bernardeau:2001qr}), we compute the power spectra and bispectrum due to clustering at tree level. We show in App.~\ref{app:PT} the perturbation theory kernels including the effect of redshift-space distortions and the bias model assumed here.

\subsection{Noise and Survey Specifications}\label{sec:experiment}
LIM experiments measure a limited volume of the Universe and can resolve temperature fluctuations up to a limited resolution. We assume the size of the voxel to be determined by the spectral and angular resolutions, with corresponding spatial dimensions along and across the line of sight
\begin{equation}
    \sigma_\parallel = \frac{c\delta\nu(1+z)}{H(z)\nu_{\rm obs}}\,,\qquad \sigma_\perp = D_{\rm M}(z)\theta_{\rm FWHM}/\sqrt{8\log 2}\,,
    \label{eq:sigma_resolution}
\end{equation}
where $\delta\nu$ is the spectral resolution, $\nu_{\rm obs}$ is the observed frequency, $D_M$ is the comoving angular diameter distance, and $\theta_{\rm FWHM}$ is the full-width half maximum of the telescope beam. We model the effect of the angular and spectral resolutions through the window $W_{\rm vox}$ introduced in Eq.~\ref{eq:obsPk} and, as in Ref.~\cite{TonyLi}, assume a Gaussian function in Fourier space. 

We consider a cylindrical survey volume aligned with the line of sight, with side $L_\parallel = c\Delta\nu(1+z)\times\left(H\nu_{\rm obs}\right)$ and base with radius $R_\perp = D_M\sqrt{\Omega_{\rm field}/\pi}$ (where $\Delta\nu$ and $\Omega_{\rm field}$ are the experimental frequency bandwidth and observed solid angle, respectively). We assume that all spatial positions within the survey are observed with the same efficiency, hence we consider the survey mask or volume window function $W_{\rm vol}$ to be a top hat in configuration space with values 1 and 0 for points within and outside this volume, respectively.

Finally, we assume Gaussian instrumental noise, such that the noise power spectrum $P_N$ is given by
\begin{equation}
    P_N = V_{\rm vox} \sigma^2_{\rm N}, \quad \text{and} \quad \sigma_{\rm N} = \frac{T_{\rm sys}}{\sqrt{N_{\rm feeds}\delta\nu t_{\rm pix}}},
\end{equation}
where $\sigma_{\rm N}$ is the standard deviation of the instrumental noise per voxel, $T_{\rm sys}$ is the system temperature of the telescope (though it can effectively include other limitations and contaminants such as continuum foregrounds~\cite{Foss_COMAP}), $N_{\rm feeds}$ is the total effective number of detectors used, and $t_{\rm pix}$ is the observing time per pixel. 

We choose to center our analysis on the COMAP~\cite{Cleary_COMAP} instrument and consider the CO(1-0) emission line observed at 29.6 GHz for a finished first phase of the experiment. Our fiducial experimental setup follows the expected sensitivities after five years of observations, which can be estimated from the early science sensitivities of COMAP~\cite{Foss_COMAP}. We consider a single 4 deg$^2$ survey area with a bandwidth of 7.7 GHz (which corresponds to the redshift range $z\in 2.4-3.4$) observed during a total time of 1000 hours (effectively accounting for $\sim 300$ hours of observation in 3 different fields for the real experiment) with 38 effective detectors (19 feeds with double polarization), with an effective system temperature of 45 K and spectral and angular resolutions of $\delta\nu=31.25$ MHz and $\theta_{\rm FWHM}=4.5$ arcmin. Finally, in order to account for the upgrade in sensitivity between the recent early science sensitivities of COMAP and the finished COMAP Y5, we use the factor $\sigma_{\rm N}^{\rm (Y5)}=\sigma_{\rm N}^{\rm (Y1)}/\sqrt{69.4}$. 

For the mean relation between luminosity and halo mass, we assume the fiducial COMAP model~\cite{Chung_COMAP}:
\begin{equation}
    \frac{L_{\rm CO}}{L_{\odot}}(M) = 4.9\times 10^{-5}\frac{C}{\left(M/M_\star\right)^A+\left(M/M_\star\right)^B}\,,
\end{equation}
and include a mean-preserving logarithmic scatter $\sigma_L$. The fiducial values adopted by COMAP for these parameters, grounded in results from Universe Machine~\cite{UM}, COLDz~\cite{COLDz} and COPSS~\cite{COPSS}, are $A=-2.85$, $B=-0.42$, $C=10^{10.63}$, $M_\star=10^{12.3}\,M_\odot$ and $\sigma_L=0.42$. This scatter results in the luminosity function extending towards very low luminosities. We neglect the contribution from line-broadening due to peculiar velocities~\cite{Chung_broadening} and, in order to ease the computations of $\mathcal{P}_1$ and other quantities depending on it, we impose an exponential cut off in the luminosity function at $20\, L_\odot$ which does not affect to the mean luminosity; we compute $\bar{n}$ including this cut off. 

\section{Position-dependent PDF}
\label{sec:PDF}
The temperature PDF shown in Eq.~\ref{eq:PT} accounts for the global matter distribution (as given in Eq.~\ref{eq:PN_trad}); however, in a finite volume of the Universe, one can expect this PDF to vary. To account for this position dependence, we begin by rewriting Eq.~\eqref{eq:PT} as a probability conditioned by the halo density perturbation field $\delta_h^{\rm v}$ smoothed over a voxel:
\begin{equation}
    \mathcal{P} (T|\delta_h^{\rm v}(\pmb{x})) = \sum_{N=0}^{\infty} \mathcal{P}_{N}(T) \mathcal{P}(N|\delta_h^{\rm v}(\pmb{x})),
\end{equation}
where the probability $ \mathcal{P}(N|\delta_h^{\rm v}(\pmb{x}))$ is a Poisson sampling with the mean determined by the density field at a given position through $N(\pmb{x}) = \bar{N}_h\left[1+\delta_h^{\rm v}(\pmb{x})\right]$ and the smoothed halo density perturbations are
\begin{equation}
    \delta_h^{\rm v}(\pmb{x}) = \int {\rm d}^3\pmb{x}'W_{\rm vox}(\pmb{x}-\pmb{x}')\delta_h(\pmb{x}'),
\end{equation}
where 
$\delta_h$ is the unsmoothed field. 

In principle, the probability $\mathcal{P} (T)$ of measuring a temperature $T$ is given by the average of $\mathcal{P} (T|\delta_h^{\rm v}(\pmb{x}))$ over realizations of the density field. However, we can write this as a spatial average by invoking the Ergodic hypothesis:
\begin{equation}
    \mathcal{P} (T) = \int {\rm d}\delta_h^{\rm v} \mathcal{P}(\delta_h^{\rm v}) \mathcal{P} (T|\delta_h^{\rm v}) = \frac{1}{V_{\text{field}}}\int_{V_{\rm field}} {\rm d}^3\pmb{x} \mathcal{P} (T|\delta_h^{\rm v}(\pmb{x}))\,. 
    \label{eq:newPT}
\end{equation}

Equation~\eqref{eq:newPT} allows us to express $\mathcal{P}(T)$ with an explicit dependence on the local overdensities, as will be required later in order to relate it with the power spectrum. However, this alternative derivation of $\mathcal{P}(T)$ is completely equivalent to the standard one discussed in the previous section, under the same set of assumptions. Let us rewrite Eq.~\eqref{eq:newPT} expanding all the terms:
\begin{equation}
    \mathcal{P} (T) = \sum_{N=0}^\infty\mathcal{P}_N(T)\int {\rm d}\delta_h^{\rm v} \mathcal{P}(\delta_h^{\rm v}) \mathcal{P} (N|\delta_h^{\rm v}) = \sum_{N=0}^\infty\mathcal{P}_N(T)\int {\rm d}\eta \mathcal{P}(\eta) \mathcal{P} (N|(\eta-\bar{N})/\bar{N})\,,
\end{equation}
where the last equality just involves the change in variable from $\delta^{\rm v}_h$ to $\eta$. Then, identifying $\mathcal{P} (N|(\eta-\bar{N}_h)/\bar{N}_h)$ as $\mathcal{P}_{\rm Poiss}(N,\eta)$, we recover Eq.~\eqref{eq:PN_trad} if we assume a lognormal distribution for $\eta$. Although we will use Eq.~\eqref{eq:PN_trad} to compute the expected $\mathcal{P}(T)$ we will stick to the expression in Eq.~\eqref{eq:newPT} to derive the covariances.


We can expand $\mathcal{P}(N|\delta_h^{\rm v}(\pmb{x}))$ for small $\delta^{\rm v}_h$ as \begin{equation}
\begin{split}
    \mathcal{P}(N|\delta_h^{\rm v}(\pmb{x})) &= \frac{(\bar{N}_h\left[1+\delta_h^{\rm v}(\pmb{x})\right])^N e^{-\bar{N}_h\left[1+\delta_h^{\rm v}(\pmb{x})\right]}}{N!} \\ &\approx \frac{\bar{N}^N_h e^{-\bar{N}_h}}{N!}\left[ 1+ \delta_h^{\rm v}(\pmb{x})(N-\bar{N}_h) + \frac{(\delta_h^{\rm v})^2(\pmb{x})}{2} \left(\bar{N}_h^2 -2\bar{N}_h N + (N-1)N \right)\right].
\end{split}
\end{equation}
Notice that in the last equality the term that factors out is the Poisson distribution $\mathcal{P}_{\text{Poiss}}(N,\bar{N}_h)$. As noted in Eq.~\ref{eq:VIDnoise}, the instrumental noise is added to the VID prediction as a convolution between the astrophysical signal and the noise PDF. This only applies to $\mathcal{P}_N(T)$, since it is the only temperature-dependent term. Therefore, the total VID to linear order in $\delta^{\rm v}_h$ is
\begin{equation}
    \mathcal{B}_{i} = \frac{N_{\text{vox}}}{V_{\text{field}}} \int\limits_{\Delta T_{i}} {\rm d}T \int\limits_{V_{\text{field}}} {\rm d}^3\pmb{x} \sum_{N=0}^{\infty} \left(\mathcal{P}_{\rm noise}*\mathcal{P}_{N}\right)(T) \mathcal{P}_{\text{Poiss}}(N,\bar{N}_h) \left[ 1+ \delta_h^{\rm v}(\pmb{x})(N-\bar{N}_h) \right]\,.
\end{equation}
We can write the spatial integral over the survey volume in the equation above as an integral over all space multiplied by the survey mask function $W_{\rm vol}(\pmb{x})$. Transforming to Fourier space, we have
\begin{equation}
    \int\limits_{V_{\rm field}} {\rm d}^3\pmb{x} \ \delta_h^{\rm v}(\pmb{x}) = \int {\rm d}^3\pmb{x}\ W_{\rm vol}(\pmb{x}) \delta_h^{\rm v}(\pmb{x})=\int \frac{{\rm d}^3\pmb{k}}{(2\pi)^3}\ W_{\rm vol}(\pmb{k}) W_{\rm vox}(-\pmb{k})\delta_h(-\pmb{k})\,,
\end{equation}
where survey window function is not normalized, so that $\int {\rm d}^3 \pmb{x} W_{\rm vol}(\pmb{x}) = V_{\rm field}$. Then, we can express the VID as:
\begin{equation}
    \mathcal{B}_{i} = N_{\text{vox}} \int\limits_{\Delta T_{i}} {\rm d}T \sum_{N=0}^{\infty} \left(\mathcal{P}_{\rm noise}*\mathcal{P}_{N}\right)(T) \mathcal{P}_{\text{Poiss}}(N,\bar{N}_h) \left[ 1+ \frac{(N-\bar{N}_h)}{V_{\text{field}}}\int \frac{d^3\pmb{k}}{(2\pi)^3} W_{\rm vol}(\pmb{k}) W_{\rm vox}(-\pmb{k})\delta_h(-\pmb{k}) \right]
    \label{eq:VIDtot}
\end{equation}

\section{Variance and Covariance}
\label{sec:covariance}
In this section we use the reformulation of the voxel temperature PDF in terms of the local halo overdensity derived in the previous section to compute the variance of the VID and its covariance with the power spectrum.

\subsection{VID}\label{sec:VID}
Given a PDF for the voxel temperatures, the {\it observed} voxel count within each temperature bin $\Delta T$ follows a multinomial distribution, since the temperature within a given voxel either lies or not in a given temperature bin and different temperature bins are mutually exclusive. The expected value for each bin $\avg{\mathcal{B}_i}=\mathcal{B}_i$, hence this is an unbiased estimator, and its variance is $\sigma^2_{{\rm bin,}i}=\mathcal{B}_i\left(1-\mathcal{B}_i/N_{\rm vox}\right)$.\footnote{The off-diagonal covariance of the VID from the sampling of the PDF is $\sigma_{{\rm bin,}ij}=-\mathcal{B}_i\mathcal{B}_j/N_{\rm vox}$. This covariance is always negative because one more voxel within a temperature bin means one less in the rest.}
Nonetheless, due to the dependence of the VID on the density field shown in the previous section, there is an additional contribution $\sigma^2_{{\rm cv}}$ to the variance of $\mathcal{B}_i$ coming from cosmic variance. In addition, different temperature bins are also correlated due to physical processes, which depend on the matter-density field,  halo bias and  line-luminosity function, correlated sky and noise structures, and processing effects; the modelling of the physical contributions to the off-diagonal covariance of the VID is beyond the scope of this study and is left for future work.

The cosmic variance only affects the astrophysical signal present in $\mathcal{P}(T)$, but it cannot be separated from the instrumental noise contributions in the VID. We therefore compute the covariance for the total VID and not for only the astrophysical contribution. 
At linear order, we have,
\begin{equation}
    \begin{split}
         \sigma^2_{{\rm cv,}i}   = \sigma^2_{\rm cv}(\mathcal{B}_{i})  & =  \avg{\mathcal{B}_{i}\mathcal{B}_i} - \avg{\mathcal{B}_{i}}\avg{\mathcal{B}_i} = \frac{N^2_{\text{vox}}}{V^2_{\text{field}}}\Upsilon_i^2\int\frac{{\rm d}^3\pmb{k}}{(2\pi)^3} W^2_{\rm vol}(\pmb{k}) W^2_{\rm vox}(-\pmb{k})P_h(k) 
         \equiv \frac{\Upsilon_i^2}{V_{\rm vox}^{2}}\sigma_{\rm vol}^2\,, 
    \end{split}
    \label{eq:VIDcosmicvar}
\end{equation}
where the integral in the last term of the first line corresponds to the variance of the halo density field on the survey volume, as indicated in the last line, and we have defined
\begin{equation}
    \Upsilon_i = \int\limits_{\Delta T_{i}} {\rm d}T \mathcal{P}_{\rm noise}*\left( \sum_{N=0}^{\infty} (N-\bar{N}_h) \mathcal{P}_{N}(T) \mathcal{P}_{\text{Poiss}}(N,\bar{N}_h)\right)\,,
\end{equation}
the sign of which depends on whether the voxel lies on a halo under or overdensity. 

We assume that these two contributions to the variance of the VID are independent, hence the total variance of $\mathcal{B}_i$ is
\begin{equation}
    \sigma^2_{{\rm tot},i} = \sigma^2_{{\rm bin,}i}+\sigma^2_{{\rm cv,}i} =
    \mathcal{B}_i\left(1-\mathcal{B}_i/N_{\rm vox}\right) + \Upsilon_i^2\sigma_{\rm vol}^2/V_{\rm vox}^{2}\,.
    \label{eq:sigmaVID_tot}
\end{equation}
Both contributions to the covariance of the VID scale generally as $V_{\rm field}$; note that  $\sigma^2_{{\rm cv}}$ only depends on the volume through $\sigma^2_{\rm vol}$, which goes roughly as $V_{\rm field}$ (modulo variations due to changes in the shape of the mask window) and that $N_{\rm vox}\propto V_{\rm field}$ in the case of $\sigma^2_{{\rm bin}}$. This may be counter intuitive, especially for the contribution coming from cosmic variance. However, note that the VID, as defined in Eq.~\eqref{eq:Balpha}, does not involve an average over positions or configurations (as it is the case of e.g., the power spectrum). If that were the case, there would be no $N_{\rm vox}$ factor in Eqs.~\eqref{eq:Balpha} and~\eqref{eq:VIDcosmicvar}, and $\sigma^2_{{\rm bin}}\propto N_{\rm vox}^{-1}$. Similarly, $\sigma^2_{{\rm cv}}$ would scale as $V_{\rm field}^{-1}$, connecting with the intuition of cosmic variance from other summary statistics. In any case, for any definition of $\mathcal{B}$, both contributions to its variance scale similarly with the volume. Finally, if we observe more than one patch on the sky, we shall divide the variance of the VID by the number $N_{\rm field}$ of fields observed (assuming all of them have the same volume and shape).


\begin{figure}[h!]
\centering
\includegraphics[width=0.49\textwidth]{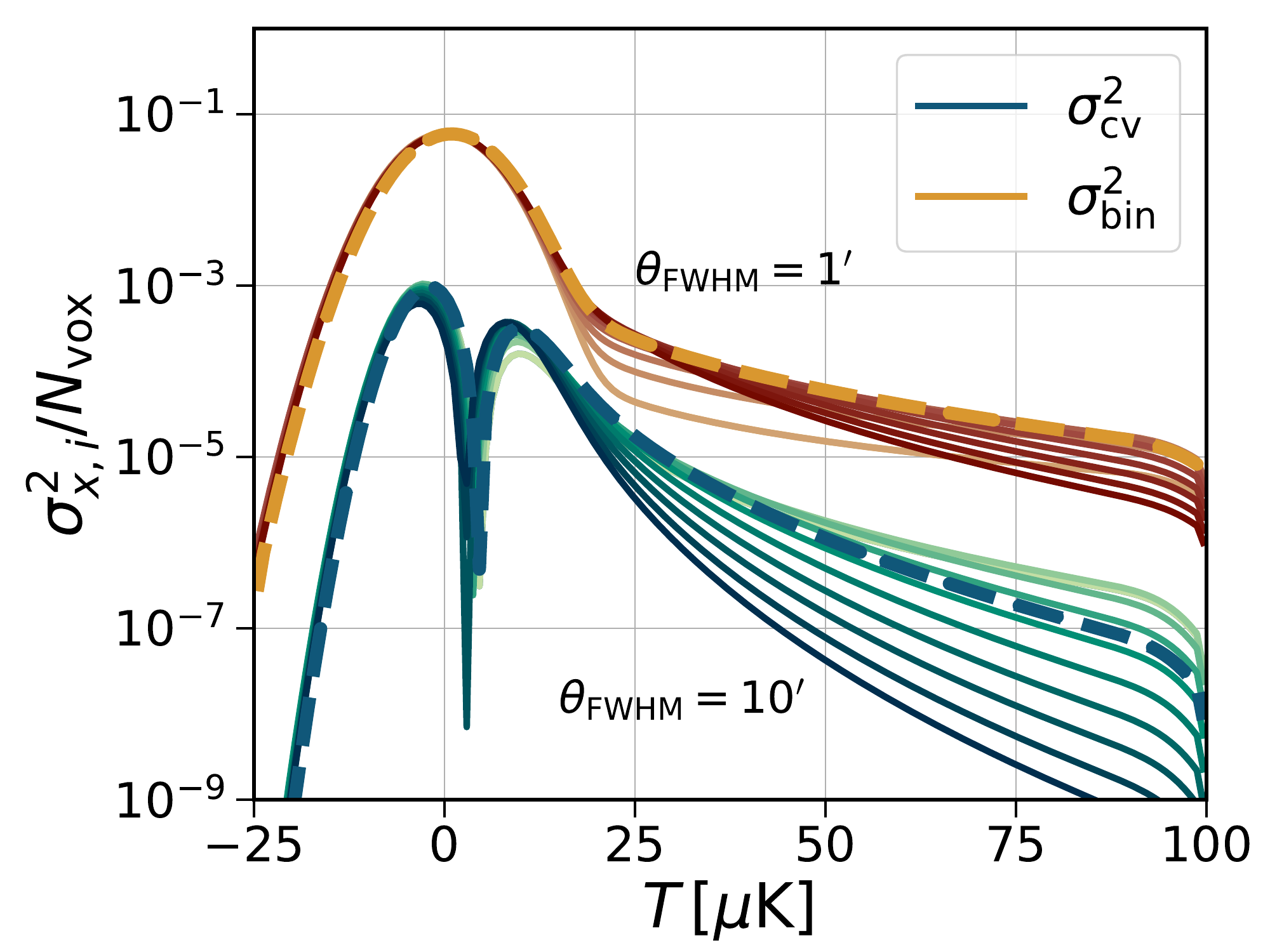}
\includegraphics[width=0.49\textwidth]{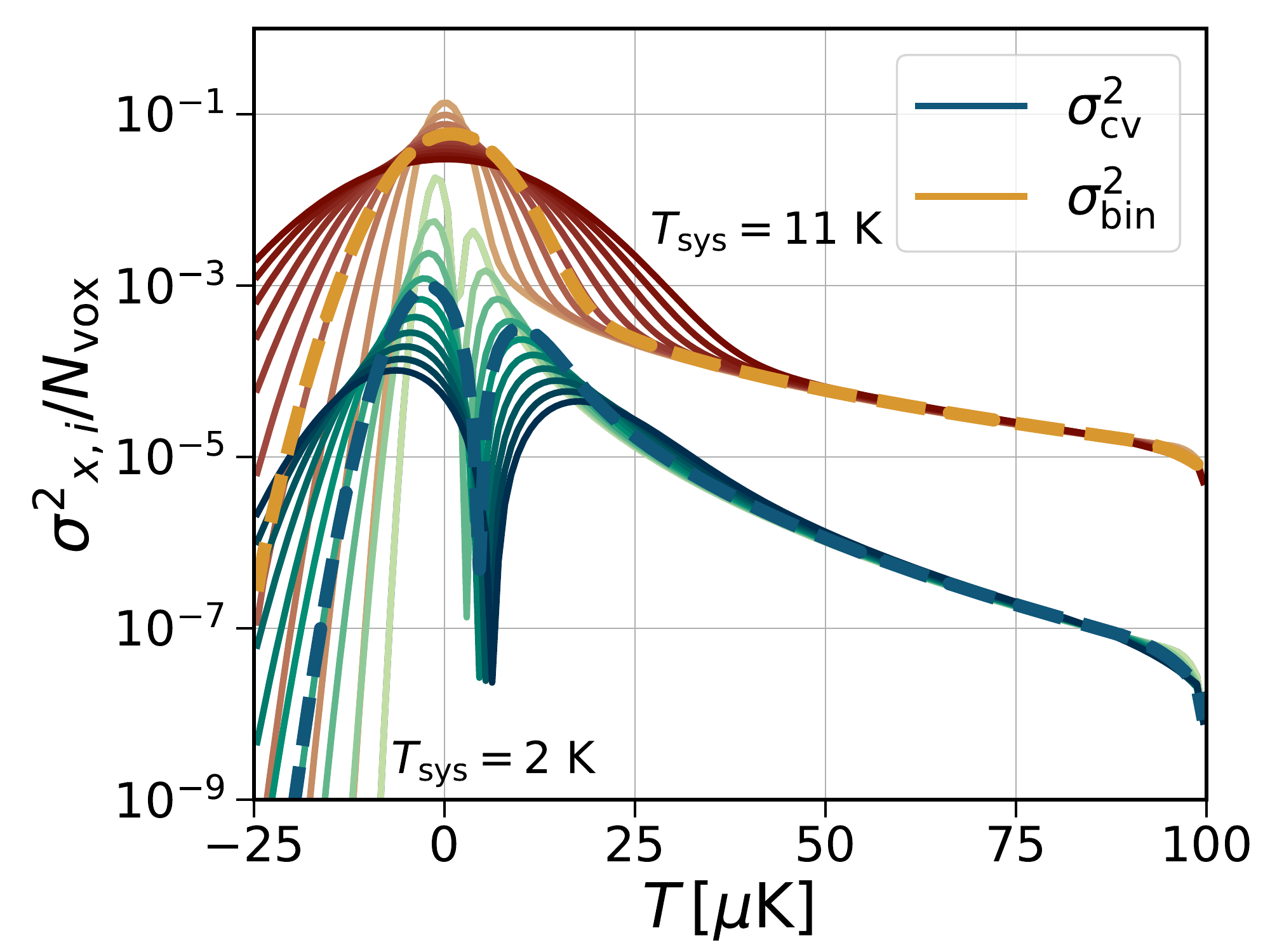}
\caption{Comparison of the bin (reds) and cosmic variance (blues) contributions to the variance of the VID for our fiducial case (thick dashed lines) and for varying the angular resolution (left) from 1' (dark) to 10' (light) while keeping every other parameter and noise per voxel fixed, and the system temperature (right) from 2 K (dark) to 11 K (light). We divide the variance by the number of voxels in each case to highlight the dependence on the resolution besides the change in the amplitude.}
\label{fig:sigma2_VID}
\end{figure}

As it is evident from Eq.~\eqref{eq:VIDcosmicvar}, $\sigma^2_{\rm cv}$ depends on the survey volume, the survey mask and the resolution, as well as $\sigma_{\rm N}$, all of which can widely vary from case to case. However, we can  compare as generally as possible the two contributions to the total variance of the VID in Eq.~\eqref{eq:sigmaVID_tot}. Their ratio is given by
\begin{equation}
    \frac{\sigma^2_{{\rm cv,}i}}{\sigma^2_{{\rm bin,}i}} = \frac{\sigma_{\rm vol}^2/V_{\rm field}}{V_{\rm vox}}\frac{\left[ \int\limits_{\Delta T_{i}} {\rm d}T \mathcal{P}_{\rm noise}*\left( \sum_{N=0}^{\infty} (N-\bar{N}_h) \mathcal{P}_{N}(T) \mathcal{P}_{\text{Poiss}}(N,\bar{N}_h)\right) \right]^2}{\int\limits_{\Delta T_{i}} {\rm d}T \mathcal{P}_{\rm noise}*\left( \sum_{N=0}^{\infty} \mathcal{P}_{N}(T) \int{\rm d}\eta \mathcal{P}_{LN}(\eta)\mathcal{P}_{\text{Poiss}}(N,\eta)\right)}\,.
    \label{eq:sigma_ratio}
\end{equation}
By definition, the two integrals above are smaller than one; in particular, the $(N-\bar{N}_h)$ term makes the integral in the numerator to be smaller for most $N$ values. Since that integral is also squared, we expect the second fraction in the right-hand side of Eq.~\eqref{eq:sigma_ratio} to be very small and dominate over the first one. Therefore, we expect the cosmic variance contribution to the variance of the VID  to be subdominant. We confirm this intuition in Fig.~\ref{fig:sigma2_VID}, where we show both contributions as function of the angular resolution (keeping all other parameters and the instrumental noise per voxel fixed) and as function of $T_{\rm sys}$, divided by the number of voxels in both cases. 

For most cases, the contribution from cosmic variance is roughly two orders of magnitude smaller than $\sigma^2_{{\rm bin}}$. This difference becomes larger at high temperatures as the angular resolution degrades, and it reduces at the temperatures where the VID starts to be dominated by the astrophysical contribution ($\sim 20\,\mu$K in our case) for better angular resolution. As we reduce $T_{\rm sys}$, the difference between the two contributions reduces at low temperatures. Therefore, this contribution cannot be neglected in cases with good angular resolution and low temperature noise.

\subsection{VID + $P(k)$}\label{sec:VID+Pk}
A joint analysis of the VID and the LIM power spectrum is key to maximize the astrophysical and cosmological information recovered from LIM experiments. In order to combine these two observables, it is necessary to model the covariance between them. One can argue that a correlation between the two signals is expected because larger temperature fluctuations result in a more extended PDF for the temperatures, and the other way around. The covariance between the VID and the power spectrum can therefore be understood as the response of the measured power spectrum to the local density perturbation and luminosity function. 

We consider the estimator for the LIM power spectrum monopole $P_0$ of a wavenumber $k_j$, which is given by
\begin{equation}
    \hat{P}_0(k_j) = \frac{1}{V_{\text{field}}} \int \limits_{V_{k_j}}\frac{{\rm d}^3 \pmb{k}}{V_{k_j}}  \tilde{\delta}_T(\pmb{k}) \tilde{\delta}_T(-\pmb{k}),
    \label{eq:PS_est}
\end{equation}
where $V_{k_j}$ is the Fourier-space volume of the $j$-th wavenumber bin, and $\tilde{\delta}_T$ is the observed brightness temperature perturbation. That is, we have that
\begin{equation}
\begin{split}
    \tilde{\delta}_T(\pmb{k}) &= \int\limits_{V_{\text{field}}} {\rm d}^3\pmb{x} e^{-i\pmb{k}\cdot\pmb{x}} \int {\rm d}^3\pmb{x}'W_{\rm vox}(\pmb{x}-\pmb{x}')\delta_h(\pmb{x}') \\
    &= \int \frac{{\rm d}^3\pmb{q}}{(2\pi)^3}  \delta_T(\pmb{k}-\pmb{q}) W_{\rm vox}(\pmb{k}-\pmb{q}) W_{\rm vol}(\pmb{q}).
\end{split}
\end{equation}

The covariance between the VID and the power spectrum is given by 
\begin{equation}
\begin{split}
    \text{Cov}[\mathcal{B}_i, \hat{P}_0(k_j)] =& \frac{N_{\text{vox}}}{V^2_{\text{field}}} 
    \Upsilon_i\int \frac{{\rm d}^2\Omega_{\hat{\pmb k}}}{4\pi} \int \frac{{\rm d}^3\pmb{q}_1}{(2\pi)^3} \int \frac{{\rm d}^3\pmb{q}_2}{(2\pi)^3} \int \frac{{\rm d}^3\pmb{q}_3}{(2\pi)^3} W_{\rm vox}(-\pmb{q}_1)W_{\rm vol}(\pmb{q}_1) W_{\rm vol}(\pmb{q}_2)\times \\ & W_{\rm vox}(\pmb{k} - \pmb{q}_2) W_{\rm vol}(\pmb{q}_3) W_{\rm vox}(-\pmb{k} - \pmb{q}_3) \avg{\delta_h(-\pmb{q}_1)\delta_T(\pmb{k}-\pmb{q}_2)\delta_T(-\pmb{k}-\pmb{q}_3)}
\end{split}
\end{equation}

From the definition of the bispectrum, we have that
\begin{equation}
    \avg{\delta_h(-\pmb{q}_1) \delta_T(\pmb{k}-\pmb{q}_2)\delta_T(-\pmb{k}-\pmb{q}_3)} = (2\pi)^3 \delta_D^3(-\pmb{q}_1-\pmb{q}_2  -\pmb{q}_3) B_{hTT}(-\pmb{q}_1, \pmb{k}-\pmb{q}_2, -\pmb{k}-\pmb{q}_3)\,,
\end{equation}
where the bispectrum above is computed according to Eq.~\ref{eq:BhTT}.
Substituting this definition into the expression for the covariance and integrating over $\pmb{q}_3$ yields
\begin{equation}
    \begin{split}
    \text{Cov}[\mathcal{B}_{i}, \hat{P}_0(k_j)] &= \frac{N_{\text{vox}}}{V^2_{\text{field}}} 
    \Upsilon_i\int \frac{{\rm d}^2\Omega_{\hat{\pmb k}}}{4\pi} \int \frac{{\rm d}^3\pmb{q}_1}{(2\pi)^3} \int \frac{{\rm d}^3\pmb{q}_2}{(2\pi)^3}W_{\rm vol}(\pmb{q}_1) W_{\rm vol}(\pmb{q}_2) W_{\rm vol}(-\pmb{q}_1 -\pmb{q}_2) \times \\ & W_{\rm vox} (-\pmb{q}_1) W_{\rm vox}(\pmb{k} - \pmb{q}_2) W_{\rm vox}(-\pmb{k}+\pmb{q}_1+\pmb{q}_2) B_{hTT}(-\pmb{q}_1, \pmb{k}-\pmb{q}_2, -\pmb{k}+\pmb{q}_1+\pmb{q}_2).
\label{eq:VID_Pk_cov}
\end{split}
\end{equation}
Assuming the theoretical model and experimental configuration described in Sec.~\ref{sec:theory}, we perform the integral shown above directly, using the \texttt{vegas} package \cite{Lepage:2020tgj}.

We define the pseudo-correlation matrix between the VID of a temperature bin $T_i$ and the power spectrum monopole of a Fourier mode $k_j$ as
\begin{equation}
    c_{ij} = \frac{{\rm Cov}\left[\mathcal{B}_i, P_0(k_j)\right]}{\sigma_{\mathcal{B}_i}\sigma_{P_0(k_j)}},
    \label{eq:corr_coeff}
\end{equation}
where the standard deviation $\sigma_{P_0(k_j)}$ is computed assuming that the Fourier modes are uncorrelated (see, e.g., Ref.~\cite{Bernal_guide}):
\begin{equation}
    \sigma^2_{P_0(k)} = \frac{1}{2}\frac{\int_{-1}^1 d\mu\  \left(\tilde{P}(k, \mu) + P_N\right)^2}{N_{\rm modes}(k)}\,, \quad \text{with} \quad N_{\rm modes}(k) = \frac{k^2 \Delta k}{4\pi^2}V_{\rm field}N_{\rm field},
\end{equation}
where $N_{\rm modes}(k)$ is the number of modes per $k$ bin of width $\Delta k$, $V_{\rm field}$ is the observed volume and $N_{\rm field}$ is the number of fields observed. The variance in the VID $\sigma_{\mathcal{B}_i}$ is calculated from Eq.~\ref{eq:sigmaVID_tot}, which includes both the Poisson noise and the effect of cosmic variance. Notice that the pseudo-correlation matrix can be greater than 1 for cases with low instrumental noise due primarily to the shortcomings of the assumed Gaussian power spectrum covariance.

\begin{figure}[t]
\centering
\includegraphics[width=\textwidth]{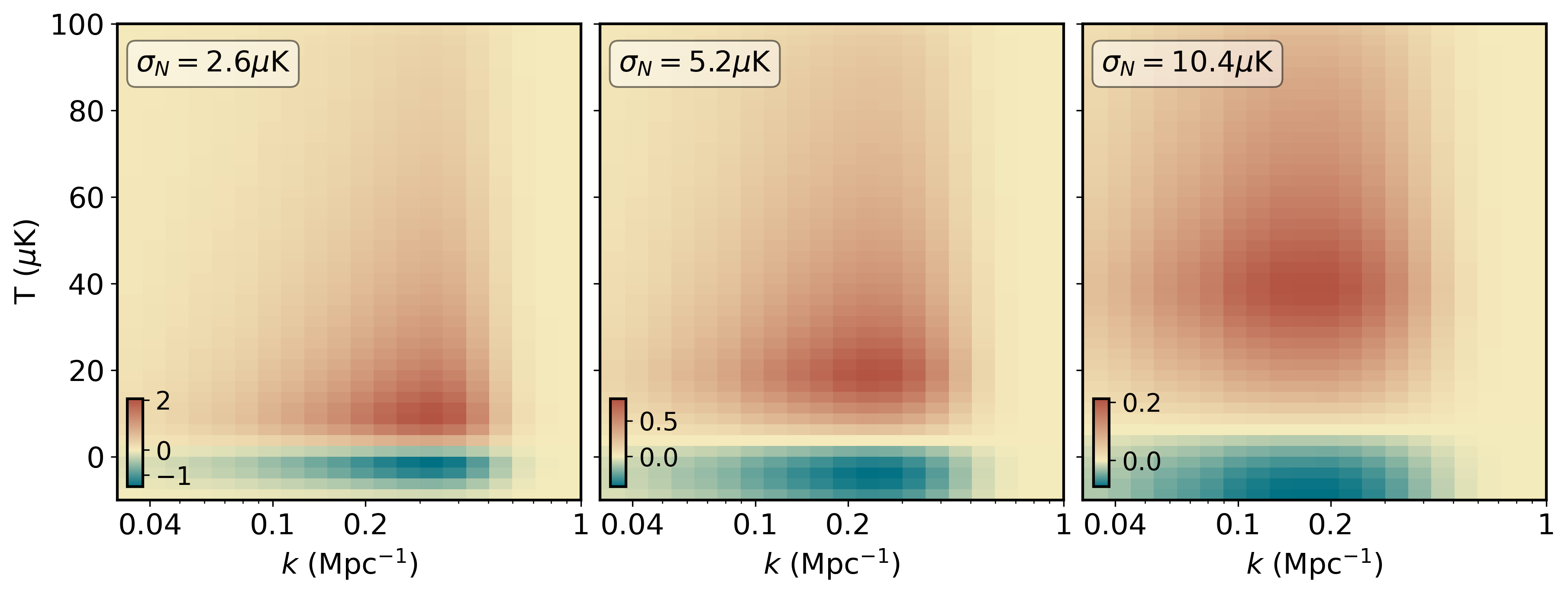}
\caption{Value of the pseudo-correlation coefficient between the VID and the power spectrum for different temperature and wavenumber bins. The middle panel corresponds to the fiducial scenario described in Sec.~\ref{sec:experiment}. The leftmost panel shows the case with half of the noise per voxel and the rightmost panel for twice the noise per voxel. Notice the change in color bar between each panel.}
\label{fig:corr}
\end{figure}

We show in Fig.~\ref{fig:corr} the expected correlation matrix between the VID and the power spectrum for each wavenumber and temperature bin. We assume the fiducial COMAP Y5 configuration described in Sec.~\ref{sec:experiment} and consider additional scenarios in which the experiment has lower or higher instrument noise. The effect this produces on the power spectrum is to increase or decrease the white noise power due to thermal noise and, on the VID, the effect is to broaden or to narrow the distribution of the noise temperature, increasing and reducing the range of temperatures for which the VID is dominated by the instrumental noise, respectively. 

Let us discuss the main qualitative features of the correlation. First, the correlation matrix is null at the smallest and largest scales probed by the power spectrum; this is due to the limited survey volume and resolution of the experiment. At low temperatures, the VID and the power spectrum are anticorrelated and become correlated as the temperature increases. This can be explained as follows. The VID at the lowest temperatures are dominated by the instrumental noise and correspond to the voxels within underdensities for which the astrophysical contribution is the dimmest. These low-temperature bins are anticorrelated with the power spectrum, since a higher power spectrum (i.e. larger temperature fluctuations) widens the astrophysical temperature PDF and the total VID --- given by the convolution between noise and astrophysical PDFs --- shifts towards higher temperatures. This results in a reduction of the values of $\mathcal{B}_i$ at low temperatures (and a corresponding increase at high temperatures, hence the positive correlation between the power spectrum and the VID). We also notice that the correlation peaks at the temperatures for which the signal-to-noise ratio for the astrophysical contribution to the VID is maximum, and reduces later at higher temperature due to a lower signal to noise. Finally, since the existence of voxels with the highest temperatures depends on the clustering at smaller scales, the correlation decays faster at large scales as the temperature increases.

The general outcome of increasing the noise per voxel is to reduce the correlation coefficient, which is to be expected since the noise is uncorrelated; note that $P_{\rm N}$ does not appear in Eq.~\eqref{eq:VID_Pk_cov}. We also notice a variety of qualitative changes in the structure of the correlation matrix. As we move from lower to higher instrumental noise, the peak of the correlation matrix shifts from smaller to larger scales. As the noise increases, the wavenumber $k$ at which the noise power spectrum crosses the signal moves towards larger scales and therefore shifts the peak of the correlation accordingly. As for the variations along the temperature bin, we can see two main effects as the instrumental noise increases: the peak of the correlation shifts towards higher temperatures, and the distribution broadens. The first feature is a consequence of the peak of the signal-to-noise ratio of the VID shifting towards higher temperatures; the second feature is a direct consequence of the broadening of the noise PDF. 

Finally, we can directly compare our results with the numerical covariance between the VID and the LIM power spectrum computed in Ref.~\cite{COMAP:2018kem}. We reproduce Fig. 2 in Ref.~\cite{COMAP:2018kem} by adopting the same set of experimental parameters and compute the correlation coefficient as described above. We present this result in App.~\ref{sec:comparison} and find good agreement between the two.

\section{Conclusions}\label{sec:conclusions}
Line-intensity mapping proposes a novel technique that provides access to large cosmological volumes and offers a complementary approach to map the large-scale structure, gleaning otherwise inaccessible astrophysical information. Due to the shared dependence on cosmology and astrophysics, the resulting line-intensity maps are very non Gaussian, hence leaving a significant amount of the information beyond the reach of power spectrum measurements. This is why the VID, which is more sensitive to the line-luminosity function, arises as a very promising summary statistic of LIM observations, especially when also combined with the power spectrum. 

We have generalized the VID formalism to account for local variations of the halo density field. This enables us to derive for the first time an analytic covariance between the LIM power spectrum and the VID --- a key step for joint analyses of one- and two-point correlations of line-intensity maps. This approach allows for a faster estimation of the covariance matrix that can be particularly useful when handling problems in which simulations are unavailable, impractical, or that require many evaluations. Furthermore, an analytical approach can offer a deeper understanding of the physical origin of such a covariance, as it clarifies the connection between the position-dependent PDFs and power spectra. We have also derived the cosmic variance contribution to the variance of the VID, and found that it is a subdominant contribution with respect to the bin variance, with possible exceptions in extreme cases.


We directly compare our results with a simulation-based analysis~\cite{COMAP:2018kem} and find that the results are consistent with each other and that the structure of the correlation matrix is successfully reproduced. While the precision of the results obtained here may be potentially limited by the modeling choices for the VID and power spectrum, the general expression for the covariance, given in Eq.~\eqref{eq:VID_Pk_cov} and the main result of this work, is general and can be refined with more accurate models. Furthermore, the correlation presented here can be easily extended to a wide range of studies of the statistical properties of intensity maps. Generalizing the covariance for higher-order multipoles of the power spectrum, for higher N-point statistics, or for different tracers of large-scale structure are natural extensions of our results.




The combination of the VID and the LIM power spectrum is expected to significantly improve constraints on theoretical parameters by LIM experiments. In particular, adding the VID to a power spectrum analysis can improve our handle on the line-luminosity function, which helps break the degeneracies between astrophysical uncertainties and cosmological features. 

\acknowledgments
We thank Marc Kamionkowski, Nickolas Kokron, Patrick Breysse and Dongwoo Chung for useful discussions. GSP was supported by the National Science Foundation Graduate Research Fellowship under Grant No.\ DGE1746891. 
JLB was supported by the Allan C. and Dorothy H. Davis Fellowship.  

\appendix

\section{Shot noise bispectrum}\label{app:shot}
In this work we consider Poissonian shot noise in all cases. Here we briefly show the derivation of the contribution to the bispectrum $B_{hTT}$. Consider an infinetisimal volume $\delta V$ that may contain $N_i=\lbrace 0,1\rbrace$ emitters, with probability of finding an emitter $\bar{n}(\pmb{x})(1+\delta_h(\pmb{x}))\delta V$. Similarly, the expected brightness temperature in that volume is $T_i=N_iX_{\rm LT}\int {\rm d}ML(M){\rm d}n/{\rm d}M\delta V/\bar{n}(\pmb{x})=\avg{T(\pmb{x})}(1+\delta_T(\pmb{x}))$, with $\delta_T=\delta T/\avg{T}$. Under these assumptions, the selfcorrelators in a given cell are:
\begin{equation}
    \avg{N_i^3} = \avg{N_i^2}=\avg{N_i} = \bar{n}(\pmb{x}_i)\delta V_i\,,\qquad \avg{T_i^n} = \avg{T^n}\,, \qquad \avg{N_iT_i} = \avg{T_i}\,,
\end{equation}
and the correlators between different cells are
\begin{equation}
\begin{split}
    \avg{n_in_j}_{i\neq j}  = & \bar{n}(\pmb{x}_i)\bar{n}(\pmb{x}_j)\delta V_i\delta V_j\left(1+\avg{\delta_h(\pmb{x}_i)\delta_h(\pmb{x}_j)}\right)\,,\\
    \avg{n_in_jn_k}_{i\neq j\neq k}  = & \bar{n}(\pmb{x}_i)\bar{n}(\pmb{x}_j)\bar{n}(\pmb{x}_k)\delta V_i\delta V_j\delta V_k\left( 1+\avg{\delta_h(\pmb{x}_i)\delta_h(\pmb{x}_j)\delta_h(\pmb{x}_k)}+ \right.\\
    &\left.\avg{\delta_h(\pmb{x}_i)\delta_h(\pmb{x}_j)}+\avg{\delta_h(\pmb{x}_i)\delta_h(\pmb{x}_k)}+\avg{\delta_h(\pmb{x}_j)\delta_h(\pmb{x}_k)}\right)\,,\\
    \avg{T_iT_j}_{i\neq j}  = & \avg{T(\pmb{x}_i)}\avg{T(\pmb{x}_j)}\delta V_i\delta V_j\left(1+\avg{\delta_T(\pmb{x}_i)\delta_T(\pmb{x}_j)}\right)\,,\\
    \avg{T_iT_jT_k}_{i\neq j\neq k}  = & \avg{T(\pmb{x}_i)}\avg{T(\pmb{x}_j)}\avg{T(\pmb{x}_k)}\delta V_i\delta V_j\delta V_k\left( 1+\avg{\delta_T(\pmb{x}_i)\delta_T(\pmb{x}_j)\delta_T(\pmb{x}_k)}+\right.\\
    &\left.\avg{\delta_T(\pmb{x}_i)\delta_T(\pmb{x}_j)} + \avg{\delta_T(\pmb{x}_i)\delta_T(\pmb{x}_k)}+\avg{\delta_T(\pmb{x}_j)\delta_T(\pmb{x}_k)}\right)\,.
\end{split}
\end{equation}
Now we can compute the multi-tracer three-point function of interest as
\begin{equation}
\begin{split}
    \int{\rm d}^3\pmb{x}_i\int{\rm d}^3\pmb{x}_j\int{\rm d}^3 & \pmb{x}_k\avg{N(\pmb{x}_i)  T(\pmb{x}_j)T(\pmb{x}_k)} = \sum_{i\neq j\neq k}  \bar{n}_i\avg{T_j}\avg{T_k}\delta V_i\delta V_j\delta V_k\left( 1+\avg{\delta_{h,i}\delta_{T,j}\delta_{T,j}}+\right.\\
     &\left. \avg{\delta_{h,i}\delta_{T,j}} + \avg{\delta_{h,i}\delta_{T,k}}+\avg{\delta_{T,j}\delta_{T,k}}\right) + \\
     \sum_{i = j\neq k} & \avg{T_i}\avg{T_k}\delta V_i\delta V_k\left( 1+\avg{\delta_{T,i}\delta_{T,j}}\right) +
     \sum_{i = k\neq j}  \avg{T_k}\avg{T_j}\delta V_k\delta V_j\left( 1+\avg{\delta_{T,k}\delta_{T,j}}\right) + \\
     \sum_{j = k\neq i} & \bar{n_i}\avg{T_j^2}\delta V_i\delta V_j\left( 1+\avg{\delta_{h,i}\delta_{T,j}}\right) + 
     \sum_{i = j = k}  \avg{T_i^2}\delta V_i\,,
\end{split}
\end{equation}
where we have converted the integrals in sums and denoted the spatial coordinates with subscripts to save space; the terms in which the sums are constrained to have the same indices correspond to Dirac deltas if we express them as integrals. We are interested in the correlation of the fluctuations:
\begin{equation}
\begin{split}
    \left\langle\left(\frac{N_i}{\bar{n}}-1\right)\left(T_j-\avg{T}\right)\left(T_k-\avg{T}\right)\right\rangle  = \frac{1}{\bar{n}}& \avg{N_iT_jT_k}  -\frac{\avg{T}}{\bar{n}}\avg{N_iT_j}  -\frac{\avg{T}}{\bar{n}}\avg{N_iT_k}-\avg{T_jT_k}+2\avg{T^2} = \\
     \avg{T}^2\avg{\delta_{h,i}\delta_{T,j}\delta_{T,k}}+\frac{\avg{T}^2}{\bar{n}} & \left[\avg{\delta_{T,i}\delta_{T,k}}\delta_D^3({\pmb{x}_i-\pmb{x}_j}) + \avg{\delta_{T,i}\delta_{T,j}}\delta_D^3({\pmb{x}_i-\pmb{x}_k})\right] + \\
     \avg{T^2}\avg{\delta_{h,i}\delta_{T,j}}\delta_D^3(\pmb{x}_j-\pmb{x}_k) & + \frac{\avg{T^2}}{\bar{n}}\delta_D^3(\pmb{x}_i-\pmb{x}_j)\delta_D^3(\pmb{x}_j-\pmb{x}_k)\,.
\end{split}
\end{equation}
Then, taking the Fourier transfrom of the expression above to compute the multi-tracer bispectrum we find
\begin{equation}
    B_{hTT}(\pmb{k}_1, \pmb{k}_2,\pmb{k}_3) = B^{\rm clust}_{hTT}(\pmb{k}_1, \pmb{k}_2,\pmb{k}_3) + \frac{P_{TT}^{\rm clust}(\pmb{k}_2)+P_{TT}^{\rm clust}(\pmb{k}_3)}{\bar{n}}+\langle T^2\rangle\left(\frac{P_{hT}^{\rm clust}(\pmb{k}_1)}{\langle T\rangle}+\frac{1}{\bar{n}}\right)\,,
\end{equation}
as expressed in Eq.~\eqref{eq:BhTT}. Similarly, the shot noise for the LIM bispectrum is given by
\begin{equation}
    B_{TTT}(\pmb{k}_1, \pmb{k}_2,\pmb{k}_3) = B^{\rm clust}_{TTT}(\pmb{k}_1, \pmb{k}_2,\pmb{k}_3) + \frac{\avg{T^2}}{\avg{T}}\left[P_{TT}^{\rm clust}(\pmb{k}_1)+P_{TT}^{\rm clust}(\pmb{k}_2)+P_{TT}^{\rm clust}(\pmb{k}_3)\right] + \avg{T^3}\,.
\end{equation}

\section{Tree-level perturbation theory kernels and biases}\label{app:PT}
The multi-tracer bispectrum required to describe the covariance between the VID and power spectrum has contributions from both clustering and shot-noise. At tree-level, the clustering term can be written as
\begin{equation}
    \begin{split}
        B_{hTT}^{\rm clust}(\pmb{k}_1, \pmb{k}_2,\pmb{k}_3) = 2\avg{T}&^2\big\{ Z^h_1(\pmb{k}_1)Z^T_1(\pmb{k}_2)Z^T_2(\pmb{k}_1, \pmb{k}_2)P_{l}(k_1)P_{l}(k_2) \\ &+ Z^h_1(\pmb{k}_1)Z^T_1(\pmb{k}_3)Z^T_2(\pmb{k}_1, \pmb{k}_3)P_{l}(k_1)P_{l}(k_3) \\ &+ Z^T_1(\pmb{k}_2)Z^T_1(\pmb{k}_3)Z^h_2(\pmb{k}_2, \pmb{k}_3)P_{l}(k_2)P_{l}(k_3)\big\},
    \end{split}
\end{equation}
where $P_l$ is the linear matter power spectrum, and $Z^{\rm X}_1$ and $Z^{\rm X}_2$ are the redshift-space kernels corresponding to tracers ${\rm X}=h,T$:
\begin{equation}
    \begin{split}
        Z_1^{\rm X}(\pmb{k}_i) \equiv& \left(b^{\rm X}_1 + f\mu_i \right)\\
        Z_2^{\rm X}(\pmb{k}_i, \pmb{k}_j) \equiv& \ b^{\rm X}_1 \left[F_2(\pmb{k}_i, \pmb{k}_j) + \frac{f\mu_{ij} k_{ij}}{2}\left(\frac{\mu_i}{k_i} + \frac{\mu_j}{k_j}\right) \right] + f\mu_{ij}^2G_2(\pmb{k}_i, \pmb{k}_j) \\ &+ \frac{f^2\mu_{ij} k_{ij}}{2} \mu_i\mu_j \left(\frac{\mu_j}{k_i} + \frac{\mu_i}{k_j}\right) + \frac{b^{\rm X}_2}{2} + \frac{b^{\rm X}_{s^2}}{2} S_2(\pmb{k}_i, \pmb{k}_j),
    \end{split}
\end{equation}
and we use $k_{ij}=|\pmb{k}_i+\pmb{k}_j|$, $\mu_{ij}=(k_i\mu_i+k_j\mu_j)/k$. $F_2$ and $G_2$ are the second-order kernels for densities and velocities, and $S_2$ is the tidal tensor kernel (see, e.g., Refs.~\cite{Bernardeau:2001qr, 2012PhRvD..86h3540B}).  Notice that the only difference between tracers in the kernels is through the bias parameters; in the case of the halo density field, the bias parameters are computed with a simple mass-average, whereas for the temperature fluctuations, they are also weighted by luminosity:
\begin{equation}
    b_\alpha^T(z) = \frac{\int {\rm d}M L(M,z) b_\alpha(M,z) \frac{dn}{dM}(M,z)}{\int {\rm d}M L(M,z) \frac{dn}{dM}(M,z)},
\end{equation}
where $\alpha$ denotes the bias parameter corresponding to either linear, quadratic, or tidal terms. We compute the halo mass function and the bias parameters $b_\alpha(M,z)$ using the Sheth-Tormen prediction \cite{Sheth:1999mn, Scoccimarro:2000gm} and assuming coevolution of halos and dark matter \cite{Chan:2012jj, 2012PhRvD..86h3540B}, so that
\begin{equation}
\begin{split}
    b_1 &= 1+\epsilon_1+E_1, \\
    b_2 &= 2\left(1-\frac{17}{21}\right)(\epsilon_1 + E_1) + \epsilon_2 + E_2, \\
    b_{s^2} &= -\frac{2}{7}\left(b_1 - 1\right),
\end{split}
\end{equation}
where
\begin{equation}
    \begin{split}
        \epsilon_1 &= \frac{\alpha\nu^2 -1}{\delta_c}, \quad \epsilon_2 = \frac{\alpha \nu^2}{\delta_c^2}\left( \alpha \nu^2 -3 \right)\\
        E_1 &= \frac{2p/\delta_c}{1+(\alpha \nu^2)^p}, \quad E_2 = E_1\left(\frac{1+2p}{\delta_c} + 2\epsilon_1\right).
    \end{split}
\end{equation}

The shot-noise term in the multi-tracer bispectrum depends on both $P_{hT}$ and $P_{TT}$, which we can write using the kernel definitions introduced above
\begin{align}
        P^{\rm{clust}}_{hT}(\pmb{k}) &= \avg{T} Z_1^h(\pmb{k})Z_1^T(\pmb{k}) P_l(k),\\
        P^{\rm{clust}}_{TT}(\pmb{k}) &= \big[\avg{T} Z_1^T(\pmb{k})\big]^2 P_l(k),\label{eq:PTTkernel}
\end{align}
where Eq.~\ref{eq:PTTkernel} is the first term in Eq.~\ref{eq:PTT}, but rewritten using the kernel $Z^T_1$.

\section{Comparison with numerical covariance}\label{sec:comparison}
Ref.~\cite{COMAP:2018kem} proposes a combined analysis of the VID and the power spectrum using a simulation-based approach. 
The covariance matrix is obtained by generating a large number of halo catalogs from ``peak patch" simulations and assigning CO luminosities based on the model outlined in Ref.~\cite{TonyLi}. We compare the results derived here with the covariance matrix measured in their work. Adopting the experimental design they define in Table 1 for the two COMAP phases, a voxel angular size defined by $\theta_{\rm FWHM}$, and the same astrophysical model, we compute the pseudo-correlation matrix using Eqs.~\ref{eq:sigmaVID_tot},~\ref{eq:VID_Pk_cov}, and~\ref{eq:corr_coeff}, defined as in Ref.~\cite{Ihle:2021fkp}.

\begin{figure}[h]
\centering
\includegraphics[width=0.8\textwidth]{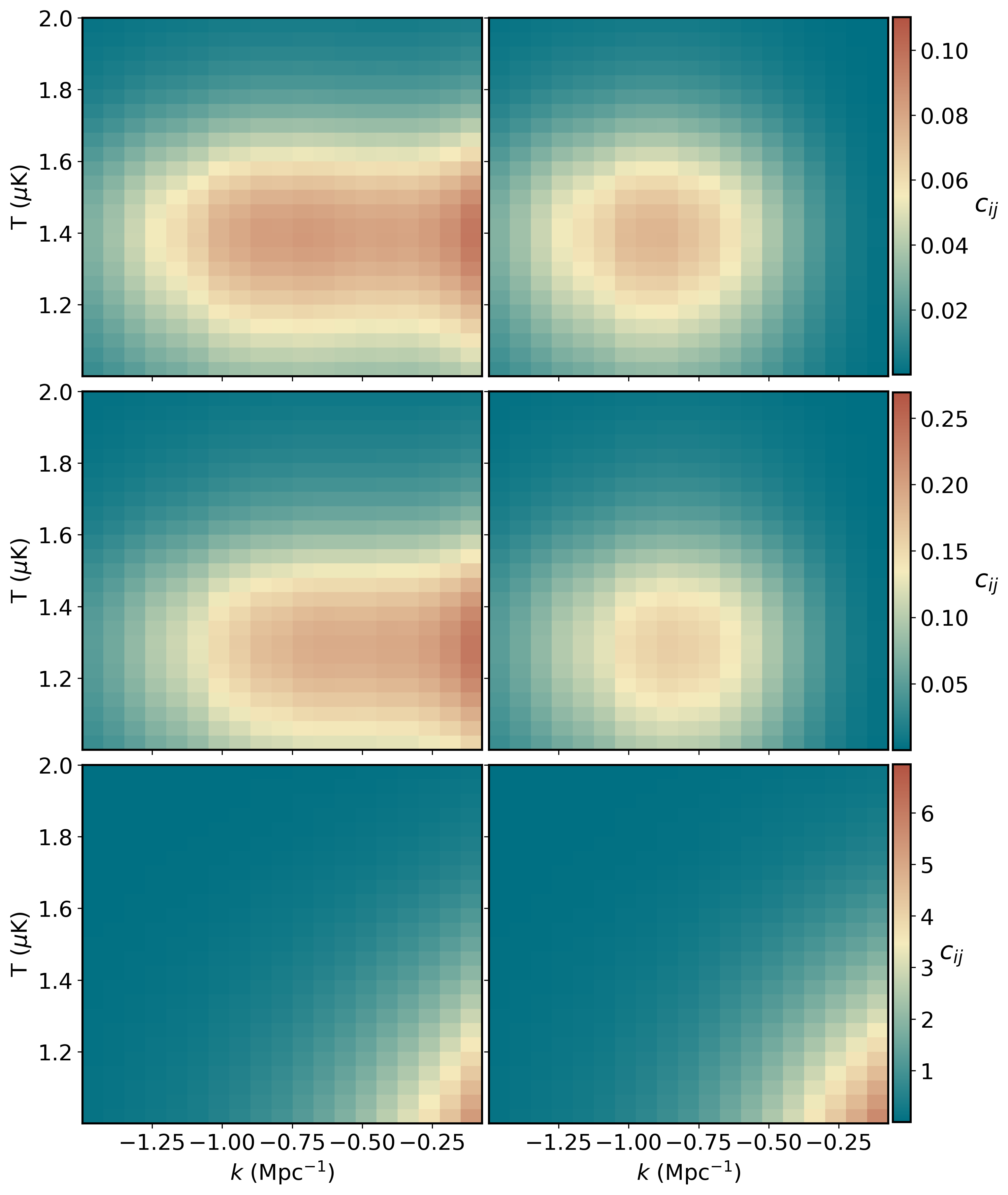}
\caption{Pseudo-correlation matrix between VID and power spectrum that can be directly compared with the simulation-based result from Ref.~\cite{COMAP:2018kem}. The three rows correspond to COMAP1, COMAP2, and signal only, and the left and right columns show the result without and with beam smoothing, respectively.}
\label{fig:corr_comparison}
\end{figure}

Fig.~\ref{fig:corr_comparison} can be directly compared to Fig. 2 in Ref.~\cite{COMAP:2018kem}, but we show only the off-diagonal block of the correlation matrix. The first row corresponds to the first stage of the COMAP experiment (COMAP1), the second row corresponds to the second phase (COMAP2), and the third row shows the result in the absence of instrumental noise. The first and second columns in Fig.~\ref{fig:corr_comparison} show the results without and with beam smoothing, respectively.\footnote{Beam smoothing has no effect on the VID measurements in our modelling of this summary statistic. This is not the case when measuring the VID from a simulated map, since the Gaussian beam extends beyond the size of the voxel.} We find good agreement between the two results and that the analytical approach outlined in this work successfully captures the important qualitative features in the correlation. 

\bibliography{ref.bib}
\bibliographystyle{utcaps}
\bigskip

\end{document}